\documentclass[reprint,prl]{revtex4-2}
\usepackage{graphicx}
\usepackage{dcolumn}
\newcommand{\brho}{\boldsymbol{\rho}}

\newcommand{\bzeta}{\boldsymbol{\zeta}}
\usepackage{bm}
\usepackage{physics}
\usepackage{amssymb}
\usepackage[english]{babel}
\usepackage{amsfonts}

\usepackage{hyperref}
\hypersetup{
    citecolor = black,
    colorlinks=true,
    linkcolor=blue,
    filecolor= black,
    urlcolor = black,
    }
\usepackage{natbib}

\usepackage{lineno}

\begin{document}

\preprint{APS/123-QED}

\title{Long-range interactions and disorder facilitate pattern formation \\ in spatial complex systems}
\author{Fabrizio Olmeda$^{1,2,\dagger}$, and Steffen Rulands$^{1,2,*}$}
\affiliation{$^{1}$Max Planck Institute for the Physics of Complex Systems, Noethnitzer Str. 38, 01187 Dresden, Germany}
\affiliation{$^{2}$Ludwigs-Maximilians-Universität München, Arnold Sommerfeld Center for Theoretical Physics, Theresienstr. 37, 80333 München, Germany}

\begin{abstract} 
Complex systems with global interactions tend to be stable if interactions between components are sufficiently homogeneous. In biological systems, which often have small copy numbers and interactions mediated by diffusing agents, noise and non-locality may affect stability. Here, we derive stability criteria for spatial complex systems with local and non-local interactions from a coarse-grained field theory with multiplicative noise. We show that long-range interactions give rise to a transition between regimes exhibiting giant density fluctuations and pattern formation. This instability is suppressed by non-reciprocity in interactions.
\end{abstract}

\maketitle

In his seminal work~\cite{May}, May studied the time evolution of a well-mixed complex system, where the rate of change of the concentration of each component depends non-linearly on all other components. 
Such systems are almost certainly stable if interactions are symmetric and the standard deviation of the Jacobian of the linearized system around such states, $\sigma$, is smaller than the inverse square root of the number of components, $L$ (May bound)~\cite{May,Wigner1958}. 
This seminal work has been extended to include non-symmetric interactions~\cite{allesina2012stability,PhysRevLett.110.168106} and dispersion in spatially extended systems, which has a destabilizing effect~\cite{baron2020dispersal}. Works on models inspired by ecosystems  considered stochasticity and showed that the stable regime can exhibit spin-glass behavior and marginally stable states~\cite{Bunin2017pre,Biroli_2018}. Global stability is also influenced by the motif structure of interaction networks~\cite{PhysRevE.98.062316,knebel2015evolutionary} and kernels \cite{pigolotti2010gaussian}.

Conditions for stability are particularly relevant in the context of biological systems. For example, in complex ecosystems, the loss of stability can lead to the extinction of species~\cite{hu2022emergent}. Further, the integrity of adult tissues relies on stable interactions between cell populations. As complex biological systems often comprise small copy numbers they exhibit strong fluctuations~\cite{schimansky2003noise}. The strength of these fluctuations typically is concentration-dependent, termed multiplicative noise. Even more than additive noise~\cite{Biancalani2017,hutt2008additive}, multiplicative noise can dominate the behavior of stochastic systems~\cite{GarcaOjalvo1999NoiseIS}. Since interactions in biological systems are often mediated by diffusing agents, they are inherently non-local. For example, in cellular systems diffusing signaling molecules give rise to interactions between cells on a length scale determined by the square root of the product of the diffusion coefficient and the degradation rate~\cite{fogler1999elements,goppel2016efficiency}.

Here, we extend these theoretical works to include these key characteristics of biological systems: non-local interactions and multiplicative noise. Specifically, we study the stability of spatially extended, stochastic complex systems with local and non-local interactions. In order to take into account the role of different noise sources, we start from a general microscopic model and derive the distribution of particle densities stemming from multiplicative noise in a field theoretical framework. Based on this, we derive conditions for stability and characterize the emergence of spatial patterns in the stable regime. We find that non-local interactions destabilize the system by facilitating pattern formation close to the boundary of stability. Multiplicative noise does not alter the conditions for stability on the mean-field level but induces giant density fluctuations in the stable regime, where pattern formation does not occur.

\begin{figure}[tb]
     \centering
\includegraphics[width=0.5\textwidth]{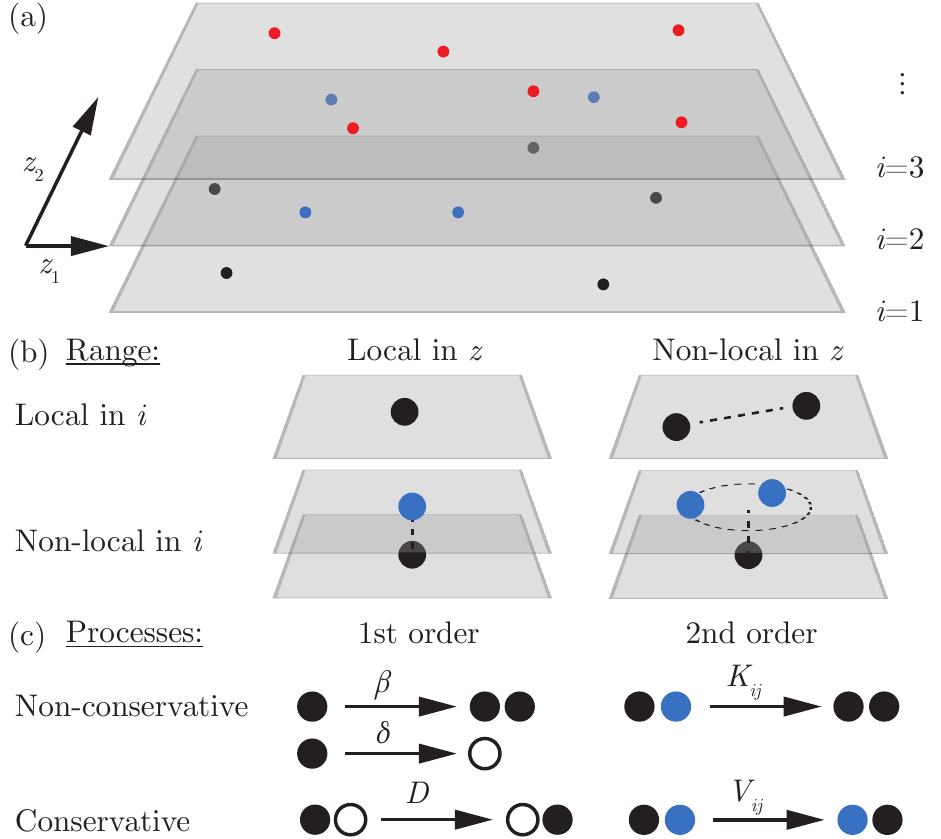}
     \caption{Schematics representing (a) the spatial and categorical degrees of freedom, (b) the possible ranges of interactions, and (c) the processes defining the microscopic model. 
     }
     \label{fig:MultiScalePlot}
\end{figure}

We consider a stochastic many-particle system of $N$ particles, where each particle, indexed by $n$, is characterized by a categorical variable $i \in \{1,\ldots,L\}$ termed component and a position $\vec{z}$ [\hyperref[fig:MultiScalePlot]{Fig.~\ref{fig:MultiScalePlot}(a)}]. In the context of ecosystems, this describes $N$ individuals belonging to $L$ species which share a habitat. 
We consider general interactions between particles that can be local or non-local in $\vec{z}$ and in $i$ [\hyperref[fig:MultiScalePlot]{Fig.~\ref{fig:MultiScalePlot}(b)}] and that may be classified based on whether they conserve global particle numbers or not [\hyperref[fig:MultiScalePlot]{Fig.~\ref{fig:MultiScalePlot}(c)}].
Within each of these classes, we consider a minimal set of microscopic rules which contribute at most quadratically to the field theory. 

Non-conservative interactions comprise a birth-death process with rates $\beta_i$ and $\delta_i$, respectively. Since interactions are non-local, these rates are functions of the positions of all other particles across components. As a second-order process, we consider Lotka-Volterra type interactions, whereupon a local or non-local interaction between a pair of particles one particle is substituted by a copy of the other. We denote the rate of such interactions between particles of components $i$ and $j$ by $K_{ij}$. In the context of ecosystems, such interactions mimic predator-prey, mutualistic, or competitive interactions between species~\cite{volterra1928variations,lotka1920analytical}. 
Conservative processes globally maintain the number of particles in each component. These comprise particle diffusion with diffusivity $D_i$ and, as a minimal process of second order, particles may move along gradients of a two-body potential $V_{ij}(|\vec{z}_i-\vec{z}_j|)$. We here consider scalar fields, such that processes like rotational diffusion~\cite{ABP_Cates_Tailleur} do not contribute to the field theory.

In order to derive a description in terms of a field theory we first define the density of particles at position $\vec{z}$ as $\rho_i(\vec{z}\,) =\sum_{n=1}^N \delta(\vec{z}_{i}^{n}-\vec{z})$. In the mean-field limit the time evolution of $\mathbf{\rho}(t)$ is then given by a differential equation of the form %
$\partial_t \boldsymbol{\rho}(\vec{z},t) =  \boldsymbol{f}[\boldsymbol{\rho}(\vec{z}',t)] \,$, where bold symbols represent vectors in $i$.
In the following, we will derive expressions for conservative and non-conservative contributions to $\boldsymbol{f}[\boldsymbol{\rho}]$ and effective noise terms from the microscopic processes defining our model.

We begin the analysis by deriving the contributions stemming from non-conservative processes, which are described in the framework of Master equations. Following standard steps~\cite{Kampen}, we expand the time evolution of the probability distribution in terms of the inverse system size and find that the non-local birth-death process contributes $\mathbf{h}\left[\brho,\vec{z}\right]\circ\brho(\vec{z}\,) + \boldsymbol{\eta}$, where $\circ$ denotes the component-wise product and $\boldsymbol{\eta}$ are multiplicative Gaussian white noise with correlations $\langle \eta_i(\vec{z},t)  \eta_j(\vec{z'},t') \rangle =(\beta_i + \delta_i) \rho_i (\vec{z},t)\delta(t-t')\delta(\vec{z}-\vec{z}\,')\delta_{i,j}$ and $h_i= \beta_i - \delta_i$~\cite{Kampen,Jorg}.
We further make the simplifying assumption that interactions decay exponentially in space with a characteristic length scale denoted by $\zeta_i$. With this, we consider a kernel $\mathbf{h}$ of the form,
\begin{equation}
    \label{eq:Expansion_h}
    \mathbf{h}[\brho,\vec{z}] \approx \mathbf{h}^{0} - \mathbf{h}^{1} \circ \int \text{d} \vec{y}\,  e^{-2|\vec{z} - \vec{y}|/\bzeta} \circ \brho(\vec{y})\, ,
\end{equation}
such that the rates $\mathbf{h}^0$ and $\mathbf{h}^1$ quantify the local and non-local excess rate of birth processes compared to death processes, respectively. With this, we can then express Eq.~\eqref{eq:Expansion_h} as a solution of a differential equation for  an auxiliary field $\phi(\vec{z},t)$~\cite{Patterns_NonLocal_Reaction_Diffusion},
\begin{equation}
 \phi_i - \zeta_i^2\laplacian  \phi_i = h^0_i  - h^1_i\zeta_i\rho_i \,.
\label{eq:resources}    
 \end{equation}
Intuitively, this equation describes the steady state of a field consumed locally by particles and subject to diffusion and degradation. 
Finally, following Ref.~\cite{mobilia2007phase} yields that non-conservative two-particle interactions between different components contribute a term $K_{ij}\rho_i\rho_j$.

In order to derive the contribution of conservative processes we express them in the form of a Langevin equation, which  
describes stochastic trajectories $\vec{z}_{\,i}^{\,n}(t)$ of individual particles, 
\begin{equation}
\label{eq:diffusion}
\partial_t \vec{z}_{\,i}^{\,n} = - \sum_{j=1}^{L} \sum_{m=1}^{N_j} \grad V_{ij}(|\vec{z}_i^{\,n}-\vec{z}_j^{\,m}|)  +  \sqrt{2D_{i}}\vec{\xi}^{\,n}_i(t) \,.
\end{equation}
$\vec{\xi}^{\,n}_i(t)$ is Gaussian white noise with correlator $\langle \vec{\xi}^{\,n}_i(t) \vec{\xi}^{\,m}_j (t')\rangle = \delta(t-t')\delta_{n,m}\delta_{i,j}$ and $N_j$ are the number of particles indexed by $j$. 
The time evolution of the density then follows from Eq.~\eqref{eq:diffusion} by following standard procedures ~\cite{Dean,Kawasaki,chavanis2019generalized}. 
Taken together, the time evolution of the density $\rho_i(\vec{z})$ follows a stochastic partial differential equation of the form
\begin{equation}
\begin{split}
\label{eq:rhophiMain}
\partial_t \rho_i &= \mathcal{L}\left[\rho_i \right]   + \sum_{j\neq i}K_{ji} \rho_j \rho_i   \\
& + \div \left[\rho_i \int \text{d}\vec{y} \, \rho_i(\vec{y}) \grad V(\vec{z} - \vec{y}) \right] + \eta_i +  \div 
\vec{\xi}_i  .
\end{split}
\end{equation}
Here, we defined the operator $\mathcal{L}\left[\rho_i \right] \equiv  \phi_i \rho_i + D_{i} \Delta \rho_i$. $\vec{\xi_i}$ is Gaussian white noise stemming from the stochastic movement of particles, Eq.~\eqref{eq:diffusion}. It has correlations $\langle \vec{\xi}_i(\vec{z},t) \vec{\xi}_j(\vec{z'},t')  \rangle = 2D_{i} \rho_i(\vec{z},t)\delta(t-t')\delta(\vec{z}-\vec{z} \,')\delta_{i,j}$.
\begin{figure*}[ht]
    \centering
\includegraphics[width=0.99 \textwidth]{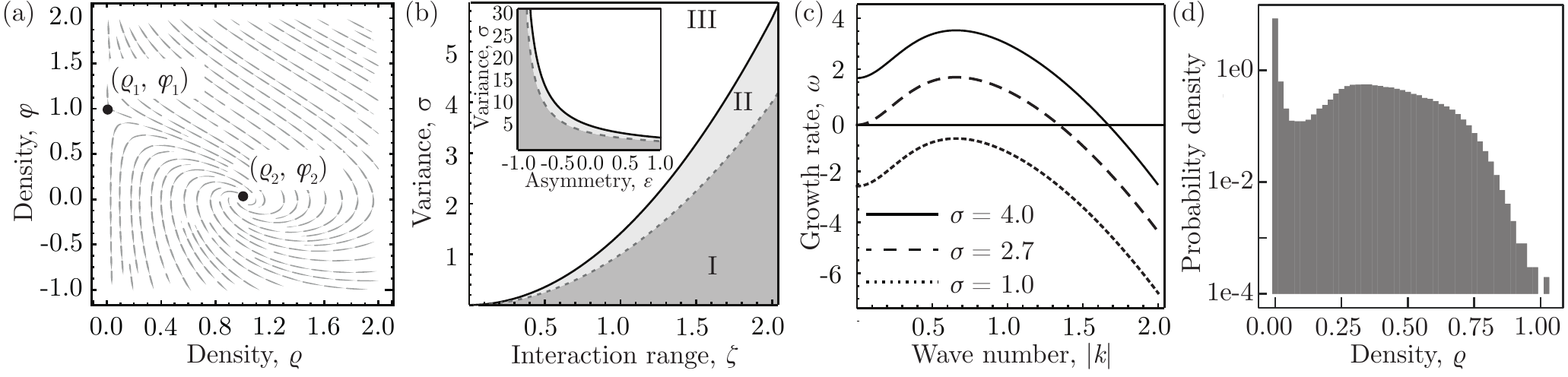}
 \caption{(a) Phase portrait of the single-component model, Eq.~\eqref{eq:rhophiMain}. Streamlines depict the time evolution of the homogeneous states $\rho$ and $\phi$. Black dots signify fixed points. (b) Phase diagram of Eq.~\eqref{eq:msrjd} as a function of the variance, range, and asymmetry (inlay) of interactions. It is valid for critical or supercritical birth-death processes, $h^0\geq 0$. The solid line depicts the condition for instability, Eq.~\eqref{eq:criticalsigma}. The dashed line represents the criterion for pattern instability. Region I exhibits giant fluctuations, II is the pattern-forming regime, and the dynamics in III exhibit unstable growth. 
 (c) Dispersion relation, Eq.~\eqref{eq:maxeigen}, in each of the three regimes of (b).
(d) Distribution of particle densities in regime I obtained by numerical solutions of  Eq.~\eqref{eq:msrjd} using the Euler-Mayorana algorithm for $d=2$ and finite central difference with integration steps $dt = 10^{-3}$ and $dx = 1$ ($L = 50$ and  $64$ sites per dimension).  Shared parameters for all panels are $h^0 =h^1\zeta=\epsilon= 1$, $D = 1$, $\zeta^2= 10$.
}
\label{fig:Fig2}
\end{figure*}

We begin our analysis of the stability conditions of Eq.~\eqref{eq:rhophiMain} by considering a system composed of a single component, $\rho$, and a constant potential $V$. We will later discuss the role of non-constant local potentials in conservative processes. Eq.~\eqref{eq:rhophiMain} then admits two spatially homogeneous stationary solutions: $\left(\rho^*_1,\phi^*_1\right) =\left(0, h^0 \right)$ is stable if the local birth-death process is subcritical, $h^0<0$, and unstable otherwise. The fixed point $\left(\rho^*_2,\phi^*_2\right) = \left( h^0/(\zeta h^1),0\right)$ exists only if $h^0 > 0$ and it is stable [\hyperref[fig:Fig2]{Fig.~\ref{fig:Fig2}(a)}]. The feedback with the field $\phi$ prevents the unbounded growth of the density even for supercritical birth-death processes.

The stability of these states with respect to spatiotemporal perturbations of the density $\rho$ can be assessed within the framework of linear stability analysis. This implies linearizing Eq.~\eqref{eq:rhophiMain} around the stationary states, $\rho(\vec{z},t)= \rho_{1,2}^* + \delta \rho(\vec{z},t)$, and studying the response of the linearized system to spatially inhomogeneous perturbations. This procedure yields a dispersion relation between the growth rate $\omega$ and the wave vector $\vec{k}$ of a spatially periodic perturbation. If the maximum of $\omega$ is positive and occurs at a finite value of $|\vec{k}|$, linear stability analysis predicts the emergence of a pattern with a finite length scale.
For single-component systems, $w$ is never positive for finite values of $|\vec{k}|$, and such systems therefore do not show pattern formation. These results are consistent with a structurally similar model that has been studied in the context of stem cells in spermatogenesis~\cite{Jorg}.

For systems comprising multiple components, the stability of stationary states may be altered by interactions between different components. In order to investigate the stability of the multi-component system, Eq.~\eqref{eq:rhophiMain}, we will first derive a phase diagram for the stability of the homogeneous stationary solutions to global perturbations. In the second step, we will then study spatiotemporal patterns in each regime. To this end, following Ref.~\cite{May}, we take $K_{ij}$ to be Gaussian distributed with mean $0$, variance $\sigma^2/L$ and covariance $\epsilon \sigma^2/L$~\footnote{With this choice, at each position $\vec{z}$ the dynamics of Eq.~\eqref{eq:rhophiMain} resembles that of a soft spin glass.}.

Following Ref.~\cite{PhysRevB.25.6860,PhysRevB.36.5388,Galla_2018} we use a
path integral representation of Eq.~\eqref{eq:rhophiMain} and perform the average over all realizations of $K_{ij}$. Within this formulation, in the limit of large $L$, we express the interaction between components in terms of a coupling to an effective response function, $\chi_i(t,t',\vec{z},\vec{z'})$, and Gaussian colored noise, $W_i$. This reduces the $2L$ coupled equations, Eq.~\eqref{eq:rhophiMain}, to $L$ uncoupled equations,
\begin{align}
\partial_t \rho_i =& \mathcal{L}\left[\rho_i \right] +  \epsilon \sigma^2 \rho_i \int_{0}^{t} \text{d}t'\int \text{d}\vec{z}' \chi_i(t,t',\vec{z},\vec{z}')\rho_i(\vec{z}',t')  \nonumber\\
&+  W_i \rho_i + \eta_i +   \div \vec{\xi_i} \, .
  \label{eq:msrjd}
\end{align}
 In Eq.~\eqref{eq:msrjd}, $W_i$ has zero mean and its correlations are proportional to the correlations of the fields ($C_{i,j}(\vec{z},\vec{z}',t,t')= \langle \rho_i(\vec{z},t) \rho_j(\vec{z}',t') \rangle$) such that  $\langle W_i(\vec{z},t) W_j(\vec{z}',t')\rangle = \sigma^2 \delta(t-t')\delta_{i,j} C_{i,j}(\vec{z},\vec{z}',t,t')$ and the response function  $\chi_i(\vec{z},\vec{z}',t,t') = \langle \partial \rho_i(\vec{z},t)/ \partial W_i(\vec{z'},t')|_{W_i = 0}\rangle $ \cite{PhysRevB.25.6860,PhysRevB.36.5388}. 
In a spatially homogeneous stationary state of Eq.~\eqref{eq:msrjd}, $\rho_i^*$, the response and correlation functions are time-independent, such that we define $\int \text{d}t' \, \chi_i(t,t') \equiv \chi_i^*$ and $C_i(t,t') = \langle \rho_i^{*2} \rangle \equiv c_i$. The stationary value of the noise is $W_i^{*}=\sqrt{\sigma^2 c_i}w$~\cite{Opper,Biscari_1995}, where $w$ is a Gaussian random variable with unitary variance and zero mean. 

In order to find stationary solutions of Eq.~\eqref{eq:msrjd} we take the mean-field limit where multiplicative noise contributions coming from density fluctuations and birth-death processes are negligible, which in our system is not expected to change the stationary states \cite{Biroli_2018}. As $W_i$ describes the effect of deterministic interactions between different components in Eq.~\eqref{eq:rhophiMain} it can not be neglected in the mean-field limit. There are two homogeneous solutions of Eq.~\eqref{eq:msrjd}, given by the random variables $(\rho_1^*, \phi_1^*) = (0,h^0 /\alpha)$ and

\begin{equation}
\label{eq:stationary_langevin}
    \rho^*_2 = 
   \frac{w\sigma \sqrt{c}+ h^0}{\Delta(\chi^*)} \Theta\left(  \ \frac{w\sigma \sqrt{c}+ h^0}{\Delta(\chi^*)}\right)
\end{equation}
where we assumed equal parameter values for all components and consequently dropped the index $i$. $\Theta(x)$ is the Heaviside step function and we defined $\Delta(\chi^*) \equiv h^1\zeta -  \epsilon \sigma^2\chi^*$. 
With $\rho^*_2$ defined the value of $\phi^*_2$ then follows from Eq.~\eqref{eq:resources}. As $w$ is a Gaussian random variable the density in the stationary state, $\rho_2^*$, follows a Gaussian distribution truncated at $0$. Therefore, as expected, the strength of particle production processes sets the mean of the stationary particle density distributions and the variance of the distribution is proportional to the variance of interactions between components. 

In order to obtain an expression for the parameters of the distribution of $\rho_2^*$, Eq.~\eqref{eq:stationary_langevin}, we derive self-consistency equations for its moments: the stationary value of the fraction of surviving components, $f_s$, the average density of particles, $M^* = \langle \rho^* \rangle$, the response function $\chi^*$, and the correlation function, $c$. 
The self-consistency equations read
\begin{equation}
\label{eq:propcorr}
\begin{split}
 f_s = \int_{-\kappa}^{\infty} \text{D}w\, ,\;
M^{*} &= \frac{\alpha \sigma \sqrt{c}}{\Delta(\chi^*)}\int_{-\kappa}^{\infty} \text{D}w \left(w+ \kappa\right)\, ,\\
\chi^{*} = \frac{\alpha }{\Delta(\chi^*) }\int_{-\kappa}^{\infty} \text{D}w ,\;
1 &= \frac{\alpha^2 \sigma^2}{\Delta(\chi^*)^2 }\int_{-\kappa}^{\infty} \text{D}w\left(w+ \kappa\right)^2  .
\end{split}
\end{equation}
Here we defined $\text{D}w \equiv\text{d}w\, e^{-w^2/2}/\sqrt{2\pi}$ and $\kappa \equiv h^0/\left(\sigma\sqrt{c}\right)$ is a threshold value of $w$ above which the density $\rho_2^*$ is a possible solution of Eq.~\eqref{eq:msrjd}. Eq.~\eqref{eq:propcorr} can be solved numerically. 

Global instability of the stationary states is associated with diverging correlations in density fluctuations, $\tilde{C} \equiv \langle \delta \rho(\vec{z},t)\delta\rho(\vec{z}\,',t') \rangle$ where $\delta \rho =\rho-\rho_2^*$~\cite{Opper, Galla_2018}. In order to determine the conditions under which correlation functions diverge, we linearize Eq.~\eqref{eq:msrjd} around the homogeneous stationary state $\rho_2^*$ and express two-point spatiotemporal correlation functions in Fourier space,
 \begin{eqnarray}
\tilde{C}(\vec{k},\vec{k}',\omega,\omega') = \frac{ \Lambda(\vec{k})\delta(\vec{k}+\vec{k}') \delta(\omega+\omega')}{ \big\langle \abs{\frac{i \omega}{\rho^*} -  \Omega(\Vec{k},\omega)}^{-2} \big\rangle_{+}^{\!-1} - f_s \sigma^2 }\, ,
  \label{eq:correlation}
\end{eqnarray}
where we defined $\Lambda(\vec{k}) =f_s(h^0/\alpha + 2Dk^2) \langle \rho^* \rangle_{+}$ and $
\Omega(\vec{k},\omega) = -\left(\frac{D}{\rho^*}\vec{k}^2+\frac{h^1\zeta }{1 +\zeta^2 \vec{k}^2}\right) +  \epsilon \sigma^2 \chi(\vec{k},\omega)\, .$
 $\langle \dots \rangle_{+}$ denotes the average over the positive fraction of surviving components $f_s$~\cite{Opper}.  

\begin{figure}[htb]
    \centering
\includegraphics[width=0.5 \textwidth]{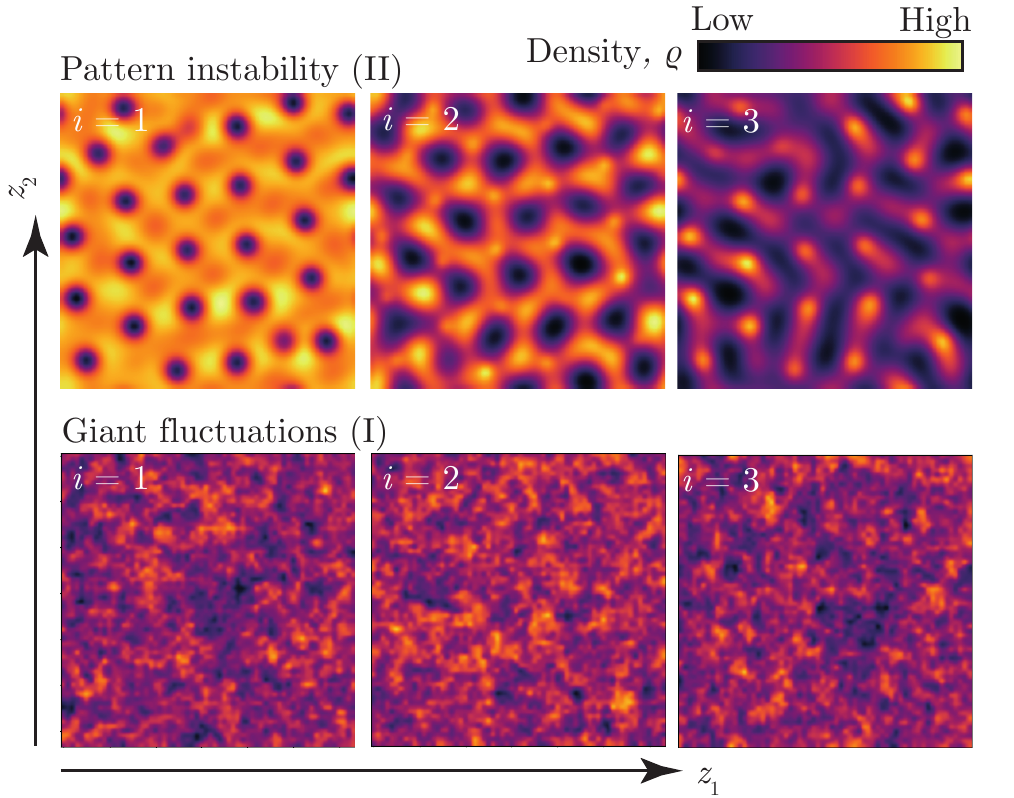}
 \caption{Snapshots of numerical simulations of Eq.~\eqref{eq:msrjd} 
 showing a pattern instability across different components (top, $\sigma = 0.65$) and giant density fluctuations (bottom, $\sigma = 0.4$). Parameters are $\zeta^2= D = 10$, $h^0 = 0$, $h^1\zeta = 1$, $L = 100$, $dt = 1e-3$, $dx =1$. Snapshots were taken at $1e5$ time steps.
}
\label{fig:Fig3}
\end{figure}

The correlation functions Eq.~\eqref{eq:correlation} of the stationary homogeneous state, $\omega\to 0$ and $|\vec{k}|\to 0$, diverge if $\Omega(0,0)^2= f_s\sigma^2$. This condition can be satisfied only if $h_0\geq 0$. 
Substituting $\Omega(0,0)^2 = f_s\sigma^2$ into Eq.~\eqref{eq:propcorr} gives the critical value of the variance of interaction strengths below which the system is stable with respect to homogeneous perturbations [dashed line in \hyperref[fig:Fig2]{Fig. 2(b)}],
\begin{equation}
\label{eq:criticalsigma}
    \sigma_c = {\sqrt{2}h^1\zeta}/{(1+\epsilon)} \, .
\end{equation}
Therefore, non-local birth-death processes globally stabilize the system. By contrast, the degree of symmetry in interactions between components, $\epsilon$, destabilizes the homogeneous state. For anti-symmetric interactions, $\epsilon = -1$, the system is always stable. Therefore, predator-prey interactions stabilize ecosystems against global perturbations. If $h^0<0$ extinction is the only stable fixed point such that interactions between components do not affect the stability to linear order.

In order to further characterize the instability we now investigate whether the stable, homogeneous solutions can be destabilized by spatial perturbations in the regime where $\sigma <\sigma_c$. For stochastic systems, pattern formation is reflected in a finite-wavelength peak of the spectral density functions in the long-term limit, $\langle |\delta\rho (\vec{k},0)|^2\rangle$ \cite{Biancalani,hernandez2004clustering}. 
 The onset of a pattern instability follows from solving $\div C(\vec{k},0)|_{\vec{k}^{*}} =0$ for $\vec{k}$ and $C(\vec{k},0)|_{\vec{k}^{*}} = 0$. Analytical solutions to this equation are feasible in the long-term limit, $\omega \rightarrow 0$, and using the approximation that averages of functions of the fields equal  the functions of averages. Following  Refs.~\cite{Opper,Roy_2019} then yields a criterion the instability of Eq.~\eqref{eq:rhophiMain} with respect to spatial perturbations. 
 
The condition for pattern instability also follows from random-matrix theory \cite{PhysRevLett.60.1895,Galla_2018}, requiring that the largest eigenvalue describing the relaxation of a perturbation of Eq.~\eqref{eq:rhophiMain} around a stable homogeneous state is positive at a finite value of $|\vec{k}|$. This yields
\begin{equation}
\label{eq:maxeigen}
 -h^1\zeta/ (1+ \zeta^2\vec{k}^2) -D\vec{k}^2/M^* + \sqrt{f_s}\sigma (1+\epsilon)>0 \,  ,
\end{equation}
and it is convex at that value [\hyperref[fig:Fig2]{Fig.~\ref{fig:Fig2}(c)}]. We find that a pattern instability is possible if $\sigma_p<\sigma<\sigma_c$ with a critical threshold $\sigma_p$ given by
$\sigma_p  =\sigma_c /\sqrt{2}$. Notably, the regime exhibiting pattern formation is fully determined by $\sigma_c$ and otherwise independent of the model parameters. $\sigma_c$ and $\sigma_p$ could be decoupled for models involving component-dependent dispersal~\cite{baron2020dispersal} or higher order interactions~\cite{Galla_2020} .

\hyperref[fig:Fig3]{Figure~\ref{fig:Fig3}} shows numerical solutions of Eq.~\eqref{eq:rhophiMain} in the pattern forming regime. Although individual components exhibit different patterns, they occur with an identical length scale. Indeed, Eq.~\eqref{eq:msrjd} implies that the dynamics of individual components are effectively coupled via a shared response function. 

 Although the system is stable against homogeneous and spatial perturbations for $\sigma<\sigma_p$ the presence of intrinsic noise can lead to the extinction of all components. In order to characterize the risk of extinction, we investigate fluctuations around the stationary state $(\rho_2^*,\phi_2^*)$ in this regime. In the limit of $|\vec{k}|\to\infty$ and $\omega\to 0$, Eq~\eqref{eq:correlation} shows an algebraic decay following $|\vec{k}|^{-2}$. This is a hallmark of giant-density fluctuations~\cite{NeutralClustering,GiantDensityFluctuations}, whose strength in a given subsystem scales stronger than predicted by the central limit theorem. These giant-density fluctuations are also reflected in finite difference simulations of Eq.~\eqref{eq:rhophiMain}, cf. \hyperref[fig:Fig2]{Fig.~\ref{fig:Fig2}(d)} and \hyperref[fig:Fig3]{Fig.~\ref{fig:Fig3}}. Therefore,  multiplicative noise  from  microscopic processes leads to strong fluctuations and an increased risk of extinction in the stable regime.

Taken together, we generalized the May bound for the stability of complex systems~\cite{May} to features characteristic of biological systems: strong, multiplicative noise stemming from small copy numbers and non-local interactions, which are often mediated by diffusing signaling factors. Starting from a microscopic model definition we derived an effective field theory. We showed that non-local interactions and multiplicative noise both alter the conditions for stability with respect to the May bound. In particular, multiplicative noise stemming from non-conservative processes gives rise to giant density fluctuations in stable stationary states whilst conservative processes could potentially destabilize the stable region of the phase space giving rise to the formation of spatial patterns. By extending the theory by non-local conservative terms or vectorial fields our work can be applied to other systems such as multi-component phase separation or active matter.

\section*{Acknowledgements}
We thank R. Mukhamadiarov and I. Di Terlizzi for helpful feedback on the manuscript. This project has received funding from the European Research Council (ERC) under the European Union’s Horizon 2020 research and innovation program (grant agreement no. 950349).
\\
\\
$^{\dagger}$ fabrizio.olmeda@ist.ac.at \\
$^*$ rulands@lmu.de

\bibliography{library}

\begin{thebibliography}{40}%
\makeatletter
\providecommand \@ifxundefined [1]{%
 \@ifx{#1\undefined}
}%
\providecommand \@ifnum [1]{%
 \ifnum #1\expandafter \@firstoftwo
 \else \expandafter \@secondoftwo
 \fi
}%
\providecommand \@ifx [1]{%
 \ifx #1\expandafter \@firstoftwo
 \else \expandafter \@secondoftwo
 \fi
}%
\providecommand \natexlab [1]{#1}%
\providecommand \enquote  [1]{``#1''}%
\providecommand \bibnamefont  [1]{#1}%
\providecommand \bibfnamefont [1]{#1}%
\providecommand \citenamefont [1]{#1}%
\providecommand \href@noop [0]{\@secondoftwo}%
\providecommand \href [0]{\begingroup \@sanitize@url \@href}%
\providecommand \@href[1]{\@@startlink{#1}\@@href}%
\providecommand \@@href[1]{\endgroup#1\@@endlink}%
\providecommand \@sanitize@url [0]{\catcode `\\12\catcode `\$12\catcode
  `\&12\catcode `\#12\catcode `\^12\catcode `\_12\catcode `\%12\relax}%
\providecommand \@@startlink[1]{}%
\providecommand \@@endlink[0]{}%
\providecommand \url  [0]{\begingroup\@sanitize@url \@url }%
\providecommand \@url [1]{\endgroup\@href {#1}{\urlprefix }}%
\providecommand \urlprefix  [0]{URL }%
\providecommand \Eprint [0]{\href }%
\providecommand \doibase [0]{https://doi.org/}%
\providecommand \selectlanguage [0]{\@gobble}%
\providecommand \bibinfo  [0]{\@secondoftwo}%
\providecommand \bibfield  [0]{\@secondoftwo}%
\providecommand \translation [1]{[#1]}%
\providecommand \BibitemOpen [0]{}%
\providecommand \bibitemStop [0]{}%
\providecommand \bibitemNoStop [0]{.\EOS\space}%
\providecommand \EOS [0]{\spacefactor3000\relax}%
\providecommand \BibitemShut  [1]{\csname bibitem#1\endcsname}%
\let\auto@bib@innerbib\@empty
\bibitem [{\citenamefont {May}(1972)}]{May}%
  \BibitemOpen
  \bibfield  {author} {\bibinfo {author} {\bibfnamefont {R.~M.}\ \bibnamefont
  {May}},\ }\bibfield  {title} {\bibinfo {title} {Will a large complex system
  be stable?},\ }\href@noop {} {\bibfield  {journal} {\bibinfo  {journal}
  {Nature}\ }\textbf {\bibinfo {volume} {238}},\ \bibinfo {pages} {413}
  (\bibinfo {year} {1972})}\BibitemShut {NoStop}%
\bibitem [{\citenamefont {Wigner}(1958)}]{Wigner1958}%
  \BibitemOpen
  \bibfield  {author} {\bibinfo {author} {\bibfnamefont {E.~P.}\ \bibnamefont
  {Wigner}},\ }\bibfield  {title} {\bibinfo {title} {On the distribution of the
  roots of certain symmetric matrices},\ }\href
  {http://www.jstor.org/stable/1970008} {\bibfield  {journal} {\bibinfo
  {journal} {Annals of Mathematics}\ }\textbf {\bibinfo {volume} {67}},\
  \bibinfo {pages} {325} (\bibinfo {year} {1958})}\BibitemShut {NoStop}%
\bibitem [{\citenamefont {Allesina}\ and\ \citenamefont
  {Tang}(2012)}]{allesina2012stability}%
  \BibitemOpen
  \bibfield  {author} {\bibinfo {author} {\bibfnamefont {S.}~\bibnamefont
  {Allesina}}\ and\ \bibinfo {author} {\bibfnamefont {S.}~\bibnamefont
  {Tang}},\ }\bibfield  {title} {\bibinfo {title} {Stability criteria for
  complex ecosystems},\ }\href@noop {} {\bibfield  {journal} {\bibinfo
  {journal} {Nature}\ }\textbf {\bibinfo {volume} {483}},\ \bibinfo {pages}
  {205} (\bibinfo {year} {2012})}\BibitemShut {NoStop}%
\bibitem [{\citenamefont {Knebel}\ \emph {et~al.}(2013)\citenamefont {Knebel},
  \citenamefont {Kr\"uger}, \citenamefont {Weber},\ and\ \citenamefont
  {Frey}}]{PhysRevLett.110.168106}%
  \BibitemOpen
  \bibfield  {author} {\bibinfo {author} {\bibfnamefont {J.}~\bibnamefont
  {Knebel}}, \bibinfo {author} {\bibfnamefont {T.}~\bibnamefont {Kr\"uger}},
  \bibinfo {author} {\bibfnamefont {M.~F.}\ \bibnamefont {Weber}},\ and\
  \bibinfo {author} {\bibfnamefont {E.}~\bibnamefont {Frey}},\ }\bibfield
  {title} {\bibinfo {title} {Coexistence and survival in conservative
  lotka-volterra networks},\ }\href
  {https://doi.org/10.1103/PhysRevLett.110.168106} {\bibfield  {journal}
  {\bibinfo  {journal} {Phys. Rev. Lett.}\ }\textbf {\bibinfo {volume} {110}},\
  \bibinfo {pages} {168106} (\bibinfo {year} {2013})}\BibitemShut {NoStop}%
\bibitem [{\citenamefont {Baron}\ and\ \citenamefont
  {Galla}(2020)}]{baron2020dispersal}%
  \BibitemOpen
  \bibfield  {author} {\bibinfo {author} {\bibfnamefont {J.~W.}\ \bibnamefont
  {Baron}}\ and\ \bibinfo {author} {\bibfnamefont {T.}~\bibnamefont {Galla}},\
  }\bibfield  {title} {\bibinfo {title} {Dispersal-induced instability in
  complex ecosystems},\ }\href@noop {} {\bibfield  {journal} {\bibinfo
  {journal} {Nature communications}\ }\textbf {\bibinfo {volume} {11}},\
  \bibinfo {pages} {1} (\bibinfo {year} {2020})}\BibitemShut {NoStop}%
\bibitem [{\citenamefont {Bunin}(2017)}]{Bunin2017pre}%
  \BibitemOpen
  \bibfield  {author} {\bibinfo {author} {\bibfnamefont {G.}~\bibnamefont
  {Bunin}},\ }\bibfield  {title} {\bibinfo {title} {Ecological communities with
  lotka-volterra dynamics},\ }\href
  {https://doi.org/10.1103/PhysRevE.95.042414} {\bibfield  {journal} {\bibinfo
  {journal} {Phys. Rev. E}\ }\textbf {\bibinfo {volume} {95}},\ \bibinfo
  {pages} {042414} (\bibinfo {year} {2017})}\BibitemShut {NoStop}%
\bibitem [{\citenamefont {Biroli}\ \emph {et~al.}(2018)\citenamefont {Biroli},
  \citenamefont {Bunin},\ and\ \citenamefont {Cammarota}}]{Biroli_2018}%
  \BibitemOpen
  \bibfield  {author} {\bibinfo {author} {\bibfnamefont {G.}~\bibnamefont
  {Biroli}}, \bibinfo {author} {\bibfnamefont {G.}~\bibnamefont {Bunin}},\ and\
  \bibinfo {author} {\bibfnamefont {C.}~\bibnamefont {Cammarota}},\ }\bibfield
  {title} {\bibinfo {title} {Marginally ﬆable equilibria in critical
  ecosyﬆems},\ }\href {https://doi.org/10.1088/1367-2630/aada58} {\bibfield
  {journal} {\bibinfo  {journal} {New Journal of Physics}\ }\textbf {\bibinfo
  {volume} {20}},\ \bibinfo {pages} {083051} (\bibinfo {year}
  {2018})}\BibitemShut {NoStop}%
\bibitem [{\citenamefont {Geiger}\ \emph {et~al.}(2018)\citenamefont {Geiger},
  \citenamefont {Knebel},\ and\ \citenamefont {Frey}}]{PhysRevE.98.062316}%
  \BibitemOpen
  \bibfield  {author} {\bibinfo {author} {\bibfnamefont {P.~M.}\ \bibnamefont
  {Geiger}}, \bibinfo {author} {\bibfnamefont {J.}~\bibnamefont {Knebel}},\
  and\ \bibinfo {author} {\bibfnamefont {E.}~\bibnamefont {Frey}},\ }\bibfield
  {title} {\bibinfo {title} {Topologically robust zero-sum games and pfaffian
  orientation: How network topology determines the long-time dynamics of the
  antisymmetric lotka-volterra equation},\ }\href
  {https://doi.org/10.1103/PhysRevE.98.062316} {\bibfield  {journal} {\bibinfo
  {journal} {Phys. Rev. E}\ }\textbf {\bibinfo {volume} {98}},\ \bibinfo
  {pages} {062316} (\bibinfo {year} {2018})}\BibitemShut {NoStop}%
\bibitem [{\citenamefont {Knebel}\ \emph {et~al.}(2015)\citenamefont {Knebel},
  \citenamefont {Weber}, \citenamefont {Kr{\"u}ger},\ and\ \citenamefont
  {Frey}}]{knebel2015evolutionary}%
  \BibitemOpen
  \bibfield  {author} {\bibinfo {author} {\bibfnamefont {J.}~\bibnamefont
  {Knebel}}, \bibinfo {author} {\bibfnamefont {M.~F.}\ \bibnamefont {Weber}},
  \bibinfo {author} {\bibfnamefont {T.}~\bibnamefont {Kr{\"u}ger}},\ and\
  \bibinfo {author} {\bibfnamefont {E.}~\bibnamefont {Frey}},\ }\bibfield
  {title} {\bibinfo {title} {Evolutionary games of condensates in coupled
  birth--death processes},\ }\href@noop {} {\bibfield  {journal} {\bibinfo
  {journal} {Nature communications}\ }\textbf {\bibinfo {volume} {6}},\
  \bibinfo {pages} {1} (\bibinfo {year} {2015})}\BibitemShut {NoStop}%
\bibitem [{\citenamefont {Pigolotti}\ \emph {et~al.}(2010)\citenamefont
  {Pigolotti}, \citenamefont {L{\'o}pez}, \citenamefont
  {Hern{\'a}ndez-Garc{\'\i}a},\ and\ \citenamefont
  {Andersen}}]{pigolotti2010gaussian}%
  \BibitemOpen
  \bibfield  {author} {\bibinfo {author} {\bibfnamefont {S.}~\bibnamefont
  {Pigolotti}}, \bibinfo {author} {\bibfnamefont {C.}~\bibnamefont
  {L{\'o}pez}}, \bibinfo {author} {\bibfnamefont {E.}~\bibnamefont
  {Hern{\'a}ndez-Garc{\'\i}a}},\ and\ \bibinfo {author} {\bibfnamefont {K.~H.}\
  \bibnamefont {Andersen}},\ }\bibfield  {title} {\bibinfo {title} {How
  gaussian competition leads to lumpy or uniform species distributions},\
  }\href@noop {} {\bibfield  {journal} {\bibinfo  {journal} {Theoretical
  Ecology}\ }\textbf {\bibinfo {volume} {3}},\ \bibinfo {pages} {89} (\bibinfo
  {year} {2010})}\BibitemShut {NoStop}%
\bibitem [{\citenamefont {Hu}\ \emph {et~al.}(2022)\citenamefont {Hu},
  \citenamefont {Amor}, \citenamefont {Barbier}, \citenamefont {Bunin},\ and\
  \citenamefont {Gore}}]{hu2022emergent}%
  \BibitemOpen
  \bibfield  {author} {\bibinfo {author} {\bibfnamefont {J.}~\bibnamefont
  {Hu}}, \bibinfo {author} {\bibfnamefont {D.~R.}\ \bibnamefont {Amor}},
  \bibinfo {author} {\bibfnamefont {M.}~\bibnamefont {Barbier}}, \bibinfo
  {author} {\bibfnamefont {G.}~\bibnamefont {Bunin}},\ and\ \bibinfo {author}
  {\bibfnamefont {J.}~\bibnamefont {Gore}},\ }\bibfield  {title} {\bibinfo
  {title} {Emergent phases of ecological diversity and dynamics mapped in
  microcosms},\ }\href@noop {} {\bibfield  {journal} {\bibinfo  {journal}
  {Science}\ }\textbf {\bibinfo {volume} {378}},\ \bibinfo {pages} {85}
  (\bibinfo {year} {2022})}\BibitemShut {NoStop}%
\bibitem [{\citenamefont {Schimansky-Geier}\ \emph {et~al.}(2003)\citenamefont
  {Schimansky-Geier}, \citenamefont {Abbott}, \citenamefont {Neiman},
  \citenamefont {Van~den Broeck} \emph {et~al.}}]{schimansky2003noise}%
  \BibitemOpen
  \bibfield  {author} {\bibinfo {author} {\bibfnamefont {L.}~\bibnamefont
  {Schimansky-Geier}}, \bibinfo {author} {\bibfnamefont {D.}~\bibnamefont
  {Abbott}}, \bibinfo {author} {\bibfnamefont {A.}~\bibnamefont {Neiman}},
  \bibinfo {author} {\bibfnamefont {C.}~\bibnamefont {Van~den Broeck}}, \emph
  {et~al.},\ }\bibfield  {title} {\bibinfo {title} {Noise in complex systems
  and stochastic dynamics}\ }(\bibinfo {organization} {SPIE},\ \bibinfo {year}
  {2003})\BibitemShut {NoStop}%
\bibitem [{\citenamefont {Biancalani}\ \emph
  {et~al.}(2017{\natexlab{a}})\citenamefont {Biancalani}, \citenamefont
  {Jafarpour},\ and\ \citenamefont {Goldenfeld}}]{Biancalani2017}%
  \BibitemOpen
  \bibfield  {author} {\bibinfo {author} {\bibfnamefont {T.}~\bibnamefont
  {Biancalani}}, \bibinfo {author} {\bibfnamefont {F.}~\bibnamefont
  {Jafarpour}},\ and\ \bibinfo {author} {\bibfnamefont {N.}~\bibnamefont
  {Goldenfeld}},\ }\bibfield  {title} {\bibinfo {title} {Giant amplification of
  noise in fluctuation-induced pattern formation},\ }\href
  {https://doi.org/10.1103/PhysRevLett.118.018101} {\bibfield  {journal}
  {\bibinfo  {journal} {Phys. Rev. Lett.}\ }\textbf {\bibinfo {volume} {118}},\
  \bibinfo {pages} {018101} (\bibinfo {year} {2017}{\natexlab{a}})}\BibitemShut
  {NoStop}%
\bibitem [{\citenamefont {Hutt}(2008)}]{hutt2008additive}%
  \BibitemOpen
  \bibfield  {author} {\bibinfo {author} {\bibfnamefont {A.}~\bibnamefont
  {Hutt}},\ }\bibfield  {title} {\bibinfo {title} {Additive noise may change
  the stability of nonlinear systems},\ }\href@noop {} {\bibfield  {journal}
  {\bibinfo  {journal} {EPL (Europhysics Letters)}\ }\textbf {\bibinfo {volume}
  {84}},\ \bibinfo {pages} {34003} (\bibinfo {year} {2008})}\BibitemShut
  {NoStop}%
\bibitem [{\citenamefont {Garc{\'i}a-Ojalvo}\ and\ \citenamefont
  {Sancho}(1999)}]{GarcaOjalvo1999NoiseIS}%
  \BibitemOpen
  \bibfield  {author} {\bibinfo {author} {\bibfnamefont {J.}~\bibnamefont
  {Garc{\'i}a-Ojalvo}}\ and\ \bibinfo {author} {\bibfnamefont {J.~M.}\
  \bibnamefont {Sancho}},\ }\bibfield  {title} {\bibinfo {title} {Noise in
  spatially extended systems}\ }(\bibinfo {year} {1999})\BibitemShut {NoStop}%
\bibitem [{\citenamefont {Fogler}\ and\ \citenamefont
  {Fogler}(1999)}]{fogler1999elements}%
  \BibitemOpen
  \bibfield  {author} {\bibinfo {author} {\bibfnamefont {H.~S.}\ \bibnamefont
  {Fogler}}\ and\ \bibinfo {author} {\bibfnamefont {S.~H.}\ \bibnamefont
  {Fogler}},\ }\href@noop {} {\emph {\bibinfo {title} {Elements of chemical
  reaction engineering}}}\ (\bibinfo  {publisher} {Pearson Educacion},\
  \bibinfo {year} {1999})\BibitemShut {NoStop}%
\bibitem [{\citenamefont {G{\"o}ppel}\ \emph {et~al.}(2016)\citenamefont
  {G{\"o}ppel}, \citenamefont {Palyulin},\ and\ \citenamefont
  {Gerland}}]{goppel2016efficiency}%
  \BibitemOpen
  \bibfield  {author} {\bibinfo {author} {\bibfnamefont {T.}~\bibnamefont
  {G{\"o}ppel}}, \bibinfo {author} {\bibfnamefont {V.~V.}\ \bibnamefont
  {Palyulin}},\ and\ \bibinfo {author} {\bibfnamefont {U.}~\bibnamefont
  {Gerland}},\ }\bibfield  {title} {\bibinfo {title} {The efficiency of driving
  chemical reactions by a physical non-equilibrium is kinetically controlled},\
  }\href@noop {} {\bibfield  {journal} {\bibinfo  {journal} {Physical Chemistry
  Chemical Physics}\ }\textbf {\bibinfo {volume} {18}},\ \bibinfo {pages}
  {20135} (\bibinfo {year} {2016})}\BibitemShut {NoStop}%
\bibitem [{\citenamefont {Volterra}(1928)}]{volterra1928variations}%
  \BibitemOpen
  \bibfield  {author} {\bibinfo {author} {\bibfnamefont {V.}~\bibnamefont
  {Volterra}},\ }\bibfield  {title} {\bibinfo {title} {Variations and
  fluctuations of the number of individuals in animal species living
  together},\ }\href@noop {} {\bibfield  {journal} {\bibinfo  {journal} {ICES
  Journal of Marine Science}\ }\textbf {\bibinfo {volume} {3}},\ \bibinfo
  {pages} {3} (\bibinfo {year} {1928})}\BibitemShut {NoStop}%
\bibitem [{\citenamefont {Lotka}(1920)}]{lotka1920analytical}%
  \BibitemOpen
  \bibfield  {author} {\bibinfo {author} {\bibfnamefont {A.~J.}\ \bibnamefont
  {Lotka}},\ }\bibfield  {title} {\bibinfo {title} {Analytical note on certain
  rhythmic relations in organic systems},\ }\href@noop {} {\bibfield  {journal}
  {\bibinfo  {journal} {Proceedings of the National Academy of Sciences}\
  }\textbf {\bibinfo {volume} {6}},\ \bibinfo {pages} {410} (\bibinfo {year}
  {1920})}\BibitemShut {NoStop}%
\bibitem [{\citenamefont {Cates}\ and\ \citenamefont
  {Tailleur}(2013)}]{ABP_Cates_Tailleur}%
  \BibitemOpen
  \bibfield  {author} {\bibinfo {author} {\bibfnamefont {M.~E.}\ \bibnamefont
  {Cates}}\ and\ \bibinfo {author} {\bibfnamefont {J.}~\bibnamefont
  {Tailleur}},\ }\bibfield  {title} {\bibinfo {title} {When are active brownian
  particles and run-and-tumble particles equivalent? consequences for
  motility-induced phase separation},\ }\href
  {https://doi.org/10.1209/0295-5075/101/20010} {\bibfield  {journal} {\bibinfo
   {journal} {{EPL} (Europhysics Letters)}\ }\textbf {\bibinfo {volume}
  {101}},\ \bibinfo {pages} {20010} (\bibinfo {year} {2013})}\BibitemShut
  {NoStop}%
\bibitem [{\citenamefont {Kampen}(2007)}]{Kampen}%
  \BibitemOpen
  \bibfield  {author} {\bibinfo {author} {\bibfnamefont {N.~V.}\ \bibnamefont
  {Kampen}},\ }\href@noop {} {\emph {\bibinfo {title} {Stochastic processes in
  physics and chemistry}}}\ (\bibinfo  {publisher} {North Holland},\ \bibinfo
  {year} {2007})\BibitemShut {NoStop}%
\bibitem [{\citenamefont {Jörg}\ \emph {et~al.}(2021)\citenamefont {Jörg},
  \citenamefont {Kitadate}, \citenamefont {Yoshida},\ and\ \citenamefont
  {Simons}}]{Jorg}%
  \BibitemOpen
  \bibfield  {author} {\bibinfo {author} {\bibfnamefont {D.~J.}\ \bibnamefont
  {Jörg}}, \bibinfo {author} {\bibfnamefont {Y.}~\bibnamefont {Kitadate}},
  \bibinfo {author} {\bibfnamefont {S.}~\bibnamefont {Yoshida}},\ and\ \bibinfo
  {author} {\bibfnamefont {B.~D.}\ \bibnamefont {Simons}},\ }\bibfield  {title}
  {\bibinfo {title} {Stem cell populations as self-renewing many-particle
  systems},\ }\href {https://doi.org/10.1146/annurev-conmatphys-041720-125707}
  {\bibfield  {journal} {\bibinfo  {journal} {Annual Review of Condensed Matter
  Physics}\ }\textbf {\bibinfo {volume} {12}},\ \bibinfo {pages} {135–153}
  (\bibinfo {year} {2021})}\BibitemShut {NoStop}%
\bibitem [{\citenamefont {Robertson}\ and\ \citenamefont
  {Skeldon}(2007)}]{Patterns_NonLocal_Reaction_Diffusion}%
  \BibitemOpen
  \bibfield  {author} {\bibinfo {author} {\bibfnamefont {N.}~\bibnamefont
  {Robertson}}\ and\ \bibinfo {author} {\bibfnamefont {A.}~\bibnamefont
  {Skeldon}},\ }\bibfield  {title} {\bibinfo {title} {Patterns in a non-local
  reaction diffusion equation}\ }(\bibinfo {year} {2007})\BibitemShut {NoStop}%
\bibitem [{\citenamefont {Mobilia}\ \emph {et~al.}(2007)\citenamefont
  {Mobilia}, \citenamefont {Georgiev},\ and\ \citenamefont
  {T{\"a}uber}}]{mobilia2007phase}%
  \BibitemOpen
  \bibfield  {author} {\bibinfo {author} {\bibfnamefont {M.}~\bibnamefont
  {Mobilia}}, \bibinfo {author} {\bibfnamefont {I.~T.}\ \bibnamefont
  {Georgiev}},\ and\ \bibinfo {author} {\bibfnamefont {U.~C.}\ \bibnamefont
  {T{\"a}uber}},\ }\bibfield  {title} {\bibinfo {title} {Phase transitions and
  spatio-temporal fluctuations in stochastic lattice lotka--volterra models},\
  }\href@noop {} {\bibfield  {journal} {\bibinfo  {journal} {Journal of
  Statistical Physics}\ }\textbf {\bibinfo {volume} {128}},\ \bibinfo {pages}
  {447} (\bibinfo {year} {2007})}\BibitemShut {NoStop}%
\bibitem [{\citenamefont {Dean}(1996)}]{Dean}%
  \BibitemOpen
  \bibfield  {author} {\bibinfo {author} {\bibfnamefont {D.~S.}\ \bibnamefont
  {Dean}},\ }\bibfield  {title} {\bibinfo {title} {Langevin equation for the
  density of a syﬆem of interacting langevin processes},\ }\href
  {https://doi.org/10.1088/0305-4470/29/24/001} {\bibfield  {journal} {\bibinfo
   {journal} {Journal of Physics A: Mathematical and General}\ }\textbf
  {\bibinfo {volume} {29}},\ \bibinfo {pages} {L613–L617} (\bibinfo {year}
  {1996})}\BibitemShut {NoStop}%
\bibitem [{\citenamefont {Kawasaki}\ and\ \citenamefont
  {Koga}(1993)}]{Kawasaki}%
  \BibitemOpen
  \bibfield  {author} {\bibinfo {author} {\bibfnamefont {K.}~\bibnamefont
  {Kawasaki}}\ and\ \bibinfo {author} {\bibfnamefont {T.}~\bibnamefont
  {Koga}},\ }\bibfield  {title} {\bibinfo {title} {Relaxation and growth of
  concentration ﬂuctuations in binary ﬂuids and polymer blends},\ }\href
  {https://doi.org/10.1016/0378-4371(93)90408-V} {\bibfield  {journal}
  {\bibinfo  {journal} {Physica A: Statiﬆical Mechanics and its
  Applications}\ }\textbf {\bibinfo {volume} {201}},\ \bibinfo {pages}
  {115–128} (\bibinfo {year} {1993})}\BibitemShut {NoStop}%
\bibitem [{\citenamefont {Chavanis}(2019)}]{chavanis2019generalized}%
  \BibitemOpen
  \bibfield  {author} {\bibinfo {author} {\bibfnamefont {P.-H.}\ \bibnamefont
  {Chavanis}},\ }\bibfield  {title} {\bibinfo {title} {The generalized
  stochastic smoluchowski equation},\ }\href@noop {} {\bibfield  {journal}
  {\bibinfo  {journal} {Entropy}\ }\textbf {\bibinfo {volume} {21}},\ \bibinfo
  {pages} {1006} (\bibinfo {year} {2019})}\BibitemShut {NoStop}%
\bibitem [{Note1()}]{Note1}%
  \BibitemOpen
  \bibinfo {note} {With this choice, at each position $\protect \vec {z}$ the
  dynamics of Eq.~\protect \eqref {eq:rhophiMain} resembles that of a soft spin
  glass.}\BibitemShut {Stop}%
\bibitem [{\citenamefont {Sompolinsky}\ and\ \citenamefont
  {Zippelius}(1982)}]{PhysRevB.25.6860}%
  \BibitemOpen
  \bibfield  {author} {\bibinfo {author} {\bibfnamefont {H.}~\bibnamefont
  {Sompolinsky}}\ and\ \bibinfo {author} {\bibfnamefont {A.}~\bibnamefont
  {Zippelius}},\ }\bibfield  {title} {\bibinfo {title} {Relaxational dynamics
  of the edwards-anderson model and the mean-field theory of spin-glasses},\
  }\href {https://doi.org/10.1103/PhysRevB.25.6860} {\bibfield  {journal}
  {\bibinfo  {journal} {Phys. Rev. B}\ }\textbf {\bibinfo {volume} {25}},\
  \bibinfo {pages} {6860} (\bibinfo {year} {1982})}\BibitemShut {NoStop}%
\bibitem [{\citenamefont {Kirkpatrick}\ and\ \citenamefont
  {Thirumalai}(1987)}]{PhysRevB.36.5388}%
  \BibitemOpen
  \bibfield  {author} {\bibinfo {author} {\bibfnamefont {T.~R.}\ \bibnamefont
  {Kirkpatrick}}\ and\ \bibinfo {author} {\bibfnamefont {D.}~\bibnamefont
  {Thirumalai}},\ }\bibfield  {title} {\bibinfo {title} {p-spin-interaction
  spin-glass models: Connections with the structural glass problem},\ }\href
  {https://doi.org/10.1103/PhysRevB.36.5388} {\bibfield  {journal} {\bibinfo
  {journal} {Phys. Rev. B}\ }\textbf {\bibinfo {volume} {36}},\ \bibinfo
  {pages} {5388} (\bibinfo {year} {1987})}\BibitemShut {NoStop}%
\bibitem [{\citenamefont {Galla}(2018)}]{Galla_2018}%
  \BibitemOpen
  \bibfield  {author} {\bibinfo {author} {\bibfnamefont {T.}~\bibnamefont
  {Galla}},\ }\bibfield  {title} {\bibinfo {title} {Dynamically evolved
  community size and stability of random lotka-volterra ecosystems (a)},\
  }\href {https://doi.org/10.1209/0295-5075/123/48004} {\bibfield  {journal}
  {\bibinfo  {journal} {{EPL} (Europhysics Letters)}\ }\textbf {\bibinfo
  {volume} {123}},\ \bibinfo {pages} {48004} (\bibinfo {year}
  {2018})}\BibitemShut {NoStop}%
\bibitem [{\citenamefont {Opper}\ and\ \citenamefont
  {Diederich}(1992)}]{Opper}%
  \BibitemOpen
  \bibfield  {author} {\bibinfo {author} {\bibfnamefont {M.}~\bibnamefont
  {Opper}}\ and\ \bibinfo {author} {\bibfnamefont {S.}~\bibnamefont
  {Diederich}},\ }\bibfield  {title} {\bibinfo {title} {Phase transition and
  1/f noise in a game dynamical model},\ }\href
  {https://doi.org/10.1103/PhysRevLett.69.1616} {\bibfield  {journal} {\bibinfo
   {journal} {Phys. Rev. Lett.}\ }\textbf {\bibinfo {volume} {69}},\ \bibinfo
  {pages} {1616} (\bibinfo {year} {1992})}\BibitemShut {NoStop}%
\bibitem [{\citenamefont {Biscari}\ and\ \citenamefont
  {Parisi}(1995)}]{Biscari_1995}%
  \BibitemOpen
  \bibfield  {author} {\bibinfo {author} {\bibfnamefont {P.}~\bibnamefont
  {Biscari}}\ and\ \bibinfo {author} {\bibfnamefont {G.}~\bibnamefont
  {Parisi}},\ }\bibfield  {title} {\bibinfo {title} {Replica symmetry breaking
  in the random replicant model},\ }\href
  {https://doi.org/10.1088/0305-4470/28/17/006} {\bibfield  {journal} {\bibinfo
   {journal} {Journal of Physics A: Mathematical and General}\ }\textbf
  {\bibinfo {volume} {28}},\ \bibinfo {pages} {4697–4708} (\bibinfo {year}
  {1995})}\BibitemShut {NoStop}%
\bibitem [{\citenamefont {Biancalani}\ \emph
  {et~al.}(2017{\natexlab{b}})\citenamefont {Biancalani}, \citenamefont
  {Jafarpour},\ and\ \citenamefont {Goldenfeld}}]{Biancalani}%
  \BibitemOpen
  \bibfield  {author} {\bibinfo {author} {\bibfnamefont {T.}~\bibnamefont
  {Biancalani}}, \bibinfo {author} {\bibfnamefont {F.}~\bibnamefont
  {Jafarpour}},\ and\ \bibinfo {author} {\bibfnamefont {N.}~\bibnamefont
  {Goldenfeld}},\ }\bibfield  {title} {\bibinfo {title} {Giant amplification of
  noise in fluctuation-induced pattern formation},\ }\href
  {https://doi.org/10.1103/PhysRevLett.118.018101} {\bibfield  {journal}
  {\bibinfo  {journal} {Phys. Rev. Lett.}\ }\textbf {\bibinfo {volume} {118}},\
  \bibinfo {pages} {018101} (\bibinfo {year} {2017}{\natexlab{b}})}\BibitemShut
  {NoStop}%
\bibitem [{\citenamefont {Hern{\'a}ndez-Garc{\'\i}a}\ and\ \citenamefont
  {L{\'o}pez}(2004)}]{hernandez2004clustering}%
  \BibitemOpen
  \bibfield  {author} {\bibinfo {author} {\bibfnamefont {E.}~\bibnamefont
  {Hern{\'a}ndez-Garc{\'\i}a}}\ and\ \bibinfo {author} {\bibfnamefont
  {C.}~\bibnamefont {L{\'o}pez}},\ }\bibfield  {title} {\bibinfo {title}
  {Clustering, advection, and patterns in a model of population dynamics with
  neighborhood-dependent rates},\ }\href@noop {} {\bibfield  {journal}
  {\bibinfo  {journal} {Physical Review E}\ }\textbf {\bibinfo {volume} {70}},\
  \bibinfo {pages} {016216} (\bibinfo {year} {2004})}\BibitemShut {NoStop}%
\bibitem [{\citenamefont {Roy}\ \emph {et~al.}(2019)\citenamefont {Roy},
  \citenamefont {Biroli}, \citenamefont {Bunin},\ and\ \citenamefont
  {Cammarota}}]{Roy_2019}%
  \BibitemOpen
  \bibfield  {author} {\bibinfo {author} {\bibfnamefont {F.}~\bibnamefont
  {Roy}}, \bibinfo {author} {\bibfnamefont {G.}~\bibnamefont {Biroli}},
  \bibinfo {author} {\bibfnamefont {G.}~\bibnamefont {Bunin}},\ and\ \bibinfo
  {author} {\bibfnamefont {C.}~\bibnamefont {Cammarota}},\ }\bibfield  {title}
  {\bibinfo {title} {Numerical implementation of dynamical mean field theory
  for disordered systems: application to the lotka–volterra model of
  ecosystems},\ }\href {https://doi.org/10.1088/1751-8121/ab1f32} {\bibfield
  {journal} {\bibinfo  {journal} {Journal of Physics A: Mathematical and
  Theoretical}\ }\textbf {\bibinfo {volume} {52}},\ \bibinfo {pages} {484001}
  (\bibinfo {year} {2019})}\BibitemShut {NoStop}%
\bibitem [{\citenamefont {Sommers}\ \emph {et~al.}(1988)\citenamefont
  {Sommers}, \citenamefont {Crisanti}, \citenamefont {Sompolinsky},\ and\
  \citenamefont {Stein}}]{PhysRevLett.60.1895}%
  \BibitemOpen
  \bibfield  {author} {\bibinfo {author} {\bibfnamefont {H.~J.}\ \bibnamefont
  {Sommers}}, \bibinfo {author} {\bibfnamefont {A.}~\bibnamefont {Crisanti}},
  \bibinfo {author} {\bibfnamefont {H.}~\bibnamefont {Sompolinsky}},\ and\
  \bibinfo {author} {\bibfnamefont {Y.}~\bibnamefont {Stein}},\ }\bibfield
  {title} {\bibinfo {title} {Spectrum of large random asymmetric matrices},\
  }\href {https://doi.org/10.1103/PhysRevLett.60.1895} {\bibfield  {journal}
  {\bibinfo  {journal} {Phys. Rev. Lett.}\ }\textbf {\bibinfo {volume} {60}},\
  \bibinfo {pages} {1895} (\bibinfo {year} {1988})}\BibitemShut {NoStop}%
\bibitem [{\citenamefont {Sidhom}\ and\ \citenamefont
  {Galla}(2020)}]{Galla_2020}%
  \BibitemOpen
  \bibfield  {author} {\bibinfo {author} {\bibfnamefont {L.}~\bibnamefont
  {Sidhom}}\ and\ \bibinfo {author} {\bibfnamefont {T.}~\bibnamefont {Galla}},\
  }\bibfield  {title} {\bibinfo {title} {Ecological communities from random
  generalized lotka-volterra dynamics with nonlinear feedback},\ }\href
  {https://doi.org/10.1103/PhysRevE.101.032101} {\bibfield  {journal} {\bibinfo
   {journal} {Phys. Rev. E}\ }\textbf {\bibinfo {volume} {101}},\ \bibinfo
  {pages} {032101} (\bibinfo {year} {2020})}\BibitemShut {NoStop}%
\bibitem [{\citenamefont {Houchmandzadeh}(2009)}]{NeutralClustering}%
  \BibitemOpen
  \bibfield  {author} {\bibinfo {author} {\bibfnamefont {B.}~\bibnamefont
  {Houchmandzadeh}},\ }\bibfield  {title} {\bibinfo {title} {Theory of neutral
  clustering for growing populations},\ }\href
  {https://doi.org/10.1103/PhysRevE.80.051920} {\bibfield  {journal} {\bibinfo
  {journal} {Phys. Rev. E}\ }\textbf {\bibinfo {volume} {80}},\ \bibinfo
  {pages} {051920} (\bibinfo {year} {2009})}\BibitemShut {NoStop}%
\bibitem [{\citenamefont {Houchmandzadeh}(2002)}]{GiantDensityFluctuations}%
  \BibitemOpen
  \bibfield  {author} {\bibinfo {author} {\bibfnamefont {B.}~\bibnamefont
  {Houchmandzadeh}},\ }\bibfield  {title} {\bibinfo {title} {Clustering of
  diffusing organisms},\ }\href {https://doi.org/10.1103/PhysRevE.66.052902}
  {\bibfield  {journal} {\bibinfo  {journal} {Phys. Rev. E}\ }\textbf {\bibinfo
  {volume} {66}},\ \bibinfo {pages} {052902} (\bibinfo {year}
  {2002})}\BibitemShut {NoStop}%
\end{thebibliography}%
\end{document}